\documentclass[prd,twocolumn,letterpaper,superscriptaddress]{revtex4}
\include{epsf}
\usepackage{graphicx}
\usepackage{pslatex}

\newcommand{\E}{\mathbf{E}}

\begin{document}
\title{Observations of the Askaryan Effect in Ice}
\author{The ANITA collaboration:
P.~W.~Gorham$^1$,
S.~W.~Barwick$^{2}$,
J.~J.~Beatty$^3$, 
D.~Z.~Besson$^4$,
W.~R.~Binns$^5$,
C.~Chen$^6$,
P.~Chen$^6$,
J.~M.~Clem$^7$,
A.~Connolly$^8$,
P.~F.~Dowkontt$^5$,
M.~A.~DuVernois$^9$, 
R.~C.~Field$^6$,
D.~Goldstein$^2$,
A.~Goodhue$^8$
C. Hast$^6$,
C.~L.~Hebert$^1$,
S.~Hoover$^8$,
M.~H.~Israel$^5$,
J.~Kowalski,$^1$
J.~G.~Learned$^1$,
K.~M.~Liewer$^{10}$,
J.~T.~Link$^{1,11}$,
E.~Lusczek$^9$,
S.~Matsuno$^{1}$,
B.~Mercurio$^3$,
C.~Miki$^{1}$,
P.~Mio\v{c}inovi\'c$^{1}$,
J.~Nam$^2$,
C.~J.~Naudet$^{10}$,
J. Ng$^6$,
R.~Nichol$^3$,
K. Palladino$^3$,
K. Reil$^6$,
A.~Romero-Wolf$^1$
M.~Rosen$^{1}$,
L.~Ruckman$^1$,
D.~Saltzberg$^8$,
D.~Seckel$^7$,
G.~S.~Varner$^{1}$,
D. Walz$^6$,
F.~Wu$^2$}
\vspace{2mm}
\noindent
\affiliation{
Dept. of Physics and Astronomy, Univ. of Hawaii, Manoa, HI 96822.   
$^2$Univ. of California, Irvine CA 92697.  
$^3$Dept. of Physics, Ohio State Univ., Columbus, OH 43210. 
$^4$Dept. of Physics and Astronomy, Univ. of Kansas, Lawrence, KS 66045. 
$^5$Dept. of Physics, Washington Univ. in St. Louis, MO 63130. 
$^6$Stanford Linear Accelerator Center, Menlo Park, CA, 94025.
$^7$University of Delaware, Newark, DE 19716. 
$^8$Dept. of Physics and Astronomy, Univ. of California, Los Angeles, CA 90095.
$^9$School of Physics and Astronomy, Univ. of Minnesota, Minneapolis, MN 55455.
$^{10}$Jet Propulsion Laboratory, Pasadena, CA 91109.
$^{11}$Currently at NASA Goddard Space Flight Center, Greenbelt, MD, 20771.
}

%\date{\today}

\begin{abstract}
We report on the first observations
of the Askaryan effect in ice: coherent impulsive radio Cherenkov radiation 
from the charge asymmetry in an electromagnetic (EM) shower.
Such radiation has
been observed in silica sand and rock salt, 
but this is the first direct observation from an EM shower in ice.
These measurements are important since
the majority of experiments to date that rely on the effect for ultra-high
energy neutrino detection are being performed using ice as the target medium.
As part of the complete validation process for the Antarctic Impulsive
Transient Antenna (ANITA) experiment, we performed an experiment
at the Stanford Linear Accelerator Center (SLAC) in June 2006
using a 7.5 metric ton ice target, yielding results fully consistent with theoretical
expectations. 
\end{abstract}
\pacs{95.55.Vj, 98.70.Sa}
%       95.55.Vj    (Neutrino etc. particle detectors in Astronomy)
%       98.70.Sa    (Cosmic Rays, including their origin, acceleration...)
%>>>>>>>>>>>>>>>>>>>>>>>>>>>>>>>>>>>>>>>>>>>>>>>>>>>>>>>>>>>>>>>>>>
%>>>>>>>>>>>>>>>>>>>>>>>>>>>>>>>>>>>>>>>>>>>>>>>>>>>>>>>>>>>>>>>>>>
%\narrowtext 

\maketitle

%\section{Introduction}

Very large scale optical Cherenkov detectors
such as the Antarctic Muon and Neutrino Detector Array (AMANDA) and
its successor IceCube have demonstrated the excellent
utility of Cherenkov radiation
in detection of neutrino interactions at $>$TeV
energies~\cite{AMANDA,IceCube} with ice as a target medium. However,
at neutrino energies above 100~PeV, the 
cubic-km scale of such detectors is inadequate to detect
more than a handful of events from the predicted cosmogenic
neutrino fluxes~\cite{Cosmonu} which represent the most compelling models
at these energies. The relevant detector volume for convincing
detection and characterization of these neutrinos is in the
range of hundreds to thousands of cubic km of water equivalent mass,
and the economic constraints of scaling
up the optical Cherenkov technique almost 
certainly preclude extending it much beyond
the size of the current IceCube detector, which will be completed
early in the next decade. 

Given the need for an alternative technique with a more
tractable economy of scale  to reach into the EeV (=1000 PeV)
energy regime, 
a new method which we denote the radio Cherenkov technique, 
has emerged within the last
decade. This method relies on properties of electromagnetic cascades 
in a dielectric medium. It was first 
hypothesized by Askaryan~\cite{Ask62} and confirmed in 2001
at SLAC ~\cite{Sal01}. High energy processes such as Compton,
Bhabha, and Moller scattering, along with positron annihilation
rapidly lead to a $\sim 20$\% negative charge asymmetry 
in the electron-photon part of a cascade.
In dense media
the shower charge bunch is compact, largely contained within a several
cm radius. At wavelengths of 10~cm or more, much larger 
than the characteristic shower bunch size, the relativistic shower bunch appears
as a single charge moving through the dielectric over a
distance of several meters or more. As an example, a typical shower
with mean Bjorken inelasticity $\langle y \rangle = 0.2$,
initiated by a $E_{\nu}=100$~PeV neutrino will create a total number of 
charged particles at shower maximum of order 
$n_{e+}+n_{e-} = \langle y \rangle E_{\nu}/1~{\rm GeV} \sim 2 \times 10^7$.
The net charge is thus $n_{e+}-n_{e-}-\sim 4\times 10^6$~e.
Since the radiated power for Cherenkov emission grows quadratically
with the charge of the emitter, the coherent power in the cm-to-m wavelength
regime is $\sim 10^{13}$ times greater than that emitted incoherently,
far exceeding any other secondary emission in optical or longer-wave
bands.

\begin{figure}[ht!]
\begin{center}
\includegraphics[width=3.35in]{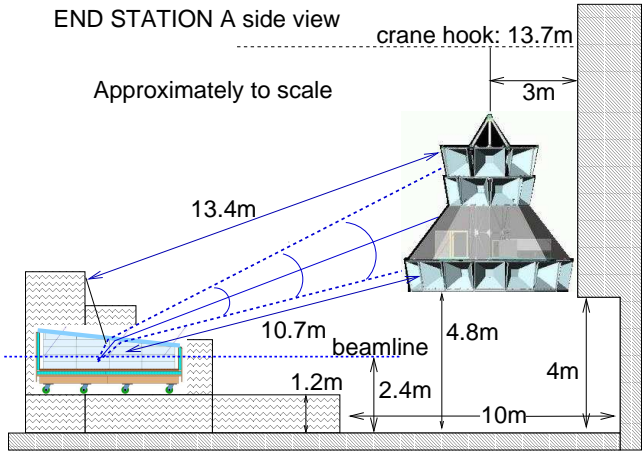} \\
\vspace{5mm}
\includegraphics[width=3.35in]{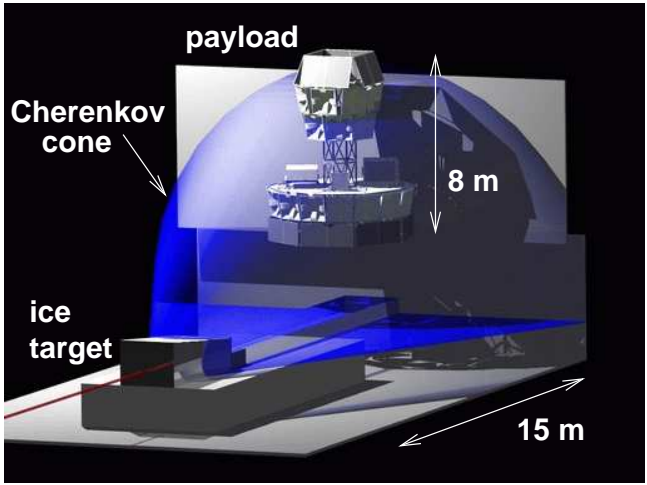} 
\caption{Top: Side view schematic of the target and receiver
arrangement in ESA. Bottom: Perspective view of the setup, showing the key elements.}
\label{ESAsetup}
\end{center}
\end{figure}

Prior to the first laboratory tests of the Askaryan effect in
1999-2000~\cite{Gor00,Sal01}, and subsequent measurements in 2002~\cite{SalSA},
it had been largely ignored since initial putative measurements of
the effect in air showers were found instead to be due to a process related
to synchrotron emission~\cite{Falcke,Suprun}. In the mid-to-late 1980's,
proposals to observe Askaryan impulses from neutrino interactions
in Antarctic ice~\cite{Gusev,Markov86,Markov60} and the Lunar regolith~\cite{Zhe88}
created a renewed interest in Askaryan's work. In the early 1990's,
the first comprehensive effort to combine EM shower simulations in ice
with electrodynamics resulted in strong support for the validity of
the methods~\cite{ZHS92}, and in the later 1990's the Radio Ice Cherenkov Experiment
(RICE)~\cite{RICE03}, and Goldstone Lunar Ultra-high energy neutrino Experiment
(GLUE)~\cite{GLUE04} began operation of experiments designed to exploit the effect.
More recently, the Fast On-orbit Recorder of Transient Events (FORTE)~\cite{FORTE04}
satellite and  the ANITA~\cite{Anitalite} experiment have extended
the method to synoptic spacecraft or balloon-payload observations of ultra-large
volumes of the Greenland or Antarctic ice sheet. 
%Plans are also underway to study
%radio extensions to the IceCube detector~\cite{AURA}, and a
%surface array on the Ross Ice Shelf~\cite{ARIANNA}.

Despite confirmation of Askaryan's theory for sand and salt, there
are important reasons to test it in ice as well, since so much study
and experimental effort have been directed at ice as the target medium.
First, although the effect is primarily determined by shower physics, the radio 
production and transmission occurs under conditions where the
properties of the medium could play a role in modifying the behavior of
the emission; the possibility of 
unknown media-dependent effects which might suppress the
emission must be explored. Second, the radio Cherenkov method is most effective at
shower energies above 10-100~PeV, where muon or other cosmic-ray backgrounds
are negligible, and the method thus ``suffers'' from the virtue of
having no natural backgrounds with which to calibrate the Cherenkov
intensity and corresponding detection efficiency. In this context, laboratory
calibrations of the radiation behavior are critical to the
accuracy of results. And finally, the increased richness of these radio
observations, which directly measure electric field strength and vector polarization,
require more comprehensive experimental treatment and validation than
observations of scalar intensity.

\begin{figure}[tb!]
\begin{center}
\vspace{2mm}
\includegraphics[width=3.4in]{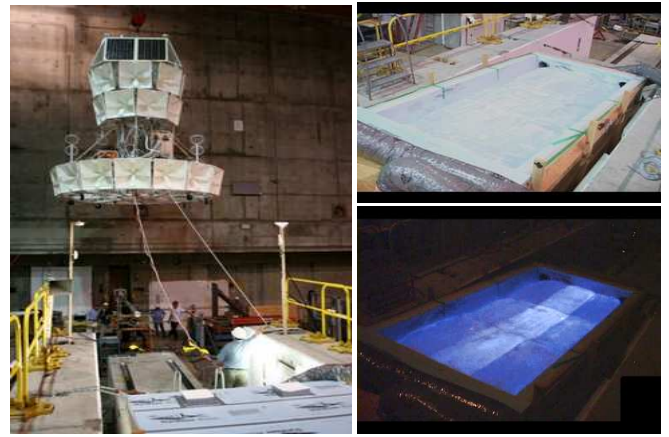}
\caption{(color online) Left: The ANITA payload (center) above and downstream of the
ice target (here covered). Right top, target with cover removed,
in ambient light. Right bottom: ice target illuminated from interior
scattered optical Cherenkov radiation.
\label{Iceshot}}
\end{center}
\end{figure}

The experiment, SLAC T486, was performed in the End Station A (ESA) facility during
the period from June 19-24, 2006. 
A target of very pure carving-grade ice
was constructed from close-packing rectangular 136~kg blocks (about 55 were used)
to form a stack approximately 2~m wide by 1.5m tall 
(at the beam entrance) by 5~m long. 
The upper surface of the ice was carved to a slope of $\sim 8^{\circ}$  
in the forward
direction giving the
block a trapezoidal longitudinal cross section along the beam axis.
This was done to avoid total-internal reflection (TIR), 
of the emerging Cherenkov radiation at the
surface. The surface after carving was measured to have a 
root-mean-square (rms) roughness of 2.3~cm. 
The beam entered this target about 40~cm above the target floor, which was lined
with 10~cm ferrite tiles to suppress reflections off the bottom. 

The showers were produced by 28.5 GeV electrons in 10 
picosecond bunches of typically $10^9$ particles.
Monte-Carlo simulations
of the showers indicate that about 90\% of the shower was contained in the 
target; the remainder was dumped into a pair of downstream concrete blocks.
In contrast to previous experiments~\cite{Sal01,SalSA}, we did not convert the electrons
to photons via a bremsstrahlung radiator. Such methods were used in
earlier Askaryan discovery experiments to avoid any initial 
excess charge in the shower development. In our case, the typical 
shower had a total composite energy of $3 \times 10^{19}$~eV, with
a total of $\sim 2 \times 10^{10}$ e$^+$e$^-$ pairs at shower maximum.
EGS simulations of the charge excess development indicate a net
charge asymmetry of about 20\%. Thus the initial electrons contribute at
most $\sim 15$\% of the total negative charge excess in the shower,
and we have corrected for this bias in the results we show here.
In addition, radio absorbing foam was in place on the front face of the 
ice, and very effectively suppressed RF signals from the upstream metal beam 
vacuum windows and air gaps.

A schematic of the experiment layout is shown in Fig.~\ref{ESAsetup}. 
The ice was contained in 
a 10~cm thick insulating foam-lined box, and a 10~cm foam lid was used during 
operation, along with a freezer unit, to maintain temperatures of between
-5 to -20 C. Such temperatures are 
adequate to avoid significant RF absorption over the several~m
pathlengths of the radiation through the ice~\cite{icepaper}.

\begin{figure}
\begin{center}
\includegraphics[width=3.6in]{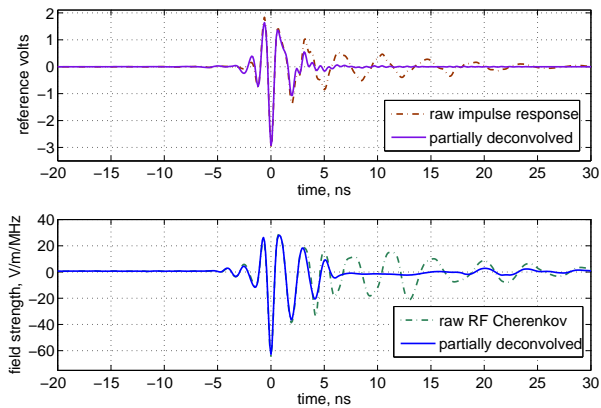}
\caption{Top: Raw, and partially-deconvolved impulse response of the ANITA
receiver system. Bottom:
Pulse received during the T486 experiment in an upper-ring antenna near
the peak of the Cherenkov cone, also showing the raw pulse, and partially
partially-deconvolved response. The apparent ``ringing'' artifact of the
raw impulses is due to group delay variation of the passband edges
of the bandpass filters employed.}
\label{waveform}
\end{center}
\end{figure}

%\begin{figure}
%\begin{center}
%\includegraphics[width=3.2in]{ALsystem05.eps}
%\includegraphics[width=3in]{T486coh.eps}
%\caption{Coherence of radio Cherenkov radiation in the T486 experiment.}
%\label{T486coh}
%\end{center}
%\end{figure}

%The coherent transition radiation (TR)~\cite{Gor00} 
%produced by the electron bunch exiting the beampipe just upstream of the target
%was suppressed by 20~cm of RF absorbing foam at the entrance to the target a
%few cm away, and additional foam helped to suppress the 
%forward TR that would be produced entering the ice, by effecting a transition
%from absorbing neoprene foam to ice, rather than air-to-ice. In addition,
%the 6$^{\circ}$  slope of the target did not begin until approximately 1~m downstream from the
%entrance, causing whatever residual TR that was produced at the entrance
%to be contained by TIR. Because of the shower multiplication, 
%TR could in any case contribute
%no more than 20\% to the total radiation power; with the additional
%measures used, we estimate that it was at most an order of magnitude
%below the Cherenkov radiation. 

The ANITA payload, consisting of an array of 32 dual-polarization quad-ridged
horn antennas was used to receive the emission at a location about 15~m away from
the center of the target, 
as shown in Fig.~\ref{Iceshot}. The antenna frequency range is from 
200-1200~MHz, which covers the majority of the frequency range over which
the RF transmissivity of ice is at its highest~\cite{icepaper}. Eight additional
vertically polarized broadband monitor antennas (four bicones and four discones) 
are used to complement the suite of horn antennas. The ANITA horn antennas
are arranged so that adjacent antennas in both the lower and upper payload sections
respond well even to a signal directed along their nearest neighbors' boresights.
This allows multiple antennas (typically 4 to 6 horns and 3 to 4 of the bicone/discones) 
to sample the arriving wavefront.
The signals are digitized by custom compact-PCI-based 8-channel 
digitizer modules~\cite{Varner}, 9 of which are used to record all 72 antenna
signals simultaneously at 2.6~Gsamples/sec. 

Figure~\ref{waveform} shows
an example of the impulse response of the system (top), and one of the
measured waveforms near the peak of the Cherenkov cone. The apparent ``ringing''
of the receiving system is due to the group delay of the
edge response of the bandpass filters, but
most of the energy arrives within a fraction of
a nanosecond, 
as determined in previous measurements of the Askaryan effect~\cite{Miocinovic}. 
In the measured T486 waveform of Fig.~\ref{waveform} (bottom), later-time
reflections from shielding and railing near the target, as well as the payload
structure, introduce some additional power into the pulse tail.

\begin{figure}
\begin{center}
\includegraphics[width=3.3in]{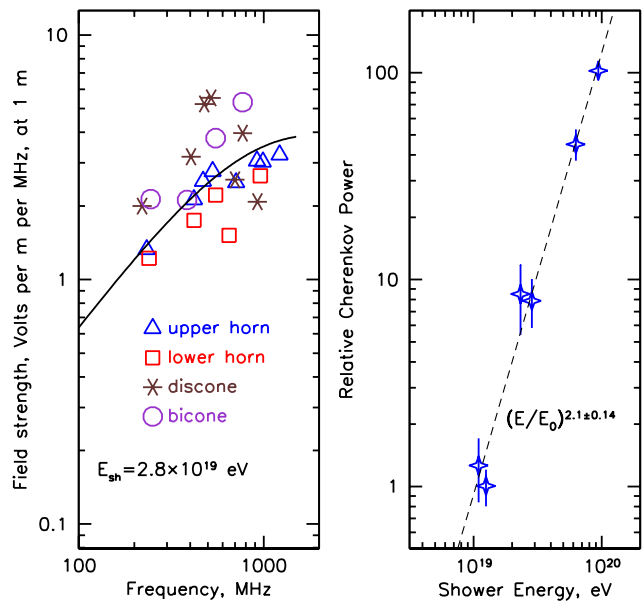}
\caption{Left: Field strength vs. frequency of radio Cherenkov radiation in the T486 experiment.
The curve is the theoretical expectation for a shower in ice at this energy.
Right: Quadratic dependence of the pulse power of the radiation detected in T486,
indicating the coherence of the Cherenkov emission.}
\label{T486field}
\end{center}
\end{figure}

In Figure~\ref{T486field} (left)
we display measurements of the absolute field strength in several different antennas,
both upper and lower quad-ridged horns, bicone, and discone antennas. 
The discone and bicone antennas have a
nearly omnidirectional response and complement the highly directive horns
by providing pulse-phase interferometry.
The uncertainty in these data are dominated by systematic, rather than
statistical errors, and are about $\pm40$\% in field strength ($\pm 3$~dB).
These are dominated by a combination of 
the 1-2dB uncertainty in the gain calibration of the antennas, 
and by comparable uncertainties in removing secondary
reflections from the measured impulse power.  
The field strengths are compared to a parameterization based
on shower+electrodynamics simulations for ice~\cite{ZHS92, Alv97}, and the agreement
is well within our experimental errors. 
Figure~\ref{T486field}(right) shows results of the scaling of the pulse power with
shower energy. The dependence is completely consistent with quadratic
scaling over the energy range we probed, indicating that the radiation
is coherent over the 200-1200~MHz frequency window. 

\begin{figure}
\begin{center}
\includegraphics[width=3.2in]{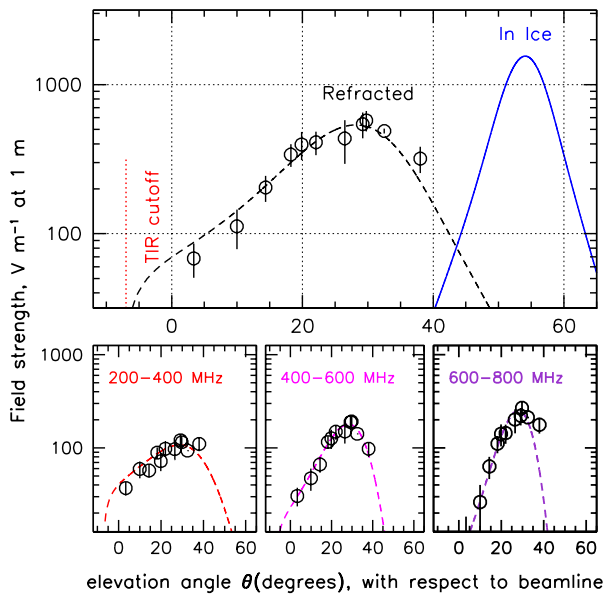}
\caption{Top: Angular dependence of the radiation for both the in-ice and refracted
case, for a frequency range from 200-800MHz, compared to data. 
The data errors are combined statistical and systematic, but with an overall
normalization that arbitrary here (but see Fig.~\ref{T486field} for
the normalization factor).
The in-ice and refracted curves are the theoretical expectation for a shower in ice at
a beam current of $10^{9}$~e$^-$per bunch and 28.5~GeV electrons, and the
refraction
includes only geometric optics.
Bottom: Same as top for three different sub-frequency bands.}
\label{T486ang}
\end{center}
\end{figure}

Figure~\ref{T486ang} shows the measured and predicted angular dependence 
of the radiation. The Cherenkov cone refracts into the forward
direction out of the ice, and is clearly delineated by the data. Here
we show statistical+systematic errors within a measurement run; 
the overall normalization (with separate systematic error) is
taken from Fig.~\ref{T486field}. We scale these data within
the overall systematic errors to match the peak of the field strength. 
The radiation frequency limit where full coherence
obtains is given approximately by the requirement that $kL\gg1$, where the 
wavenumber $k=2\pi n\nu/c$ for frequency $\nu$ and index of refraction $n$.
In this regime, the Cherenkov field strength (${\rm V~m^{-1}~Hz^{-1}}$)
can be approximated as~\cite{FORTE04}:
%\begin{equation}
$|R\E(\nu)| ~=\sqrt{2\pi}~\mu\mu_0 Q L \nu
 \sin\theta ~\exp{[-(kL)^2(\cos\theta-1/n)^2/2]}
\label{eq:cher}$
%\end{equation}
where for typical dielectrics $\mu=1$, $\mu_0= 4\pi \times 10^{-7}$ is
the permeability of free space, $L$ is the parameter determined from
the Gaussian fit of $q(z)=Q~\exp(-(z-z_{max})^2/2L^2)$ to the shower profile
with maximum at $z_{max}$,
$\theta$ is the polar angle around the shower axis, and $R$ is the distance to
the shower. For T486, $L\sim 1.2$~m.
The measured
angular dependence thus follows closely the expectations for Cherenkov radiation,
including the narrowing of the Cherenkov cone with higher frequencies. These
results further strengthen the identification of its origin.
We also measured the vector E-field polarization of the impulses and found
it to be entirely consistent with 100\% linear polarization in the plane 
containing the Poynting vector and shower momentum vectors, again completely
consistent with radio Cherenkov theory. 

%We note that the ice surface had an rms roughness approaching 1/8 wave at
%our highest frequencies (and nearly 1/4 wave for the fields propagating
%in the ice where the index of refraction is $n=1.78$). At this level, the
%roughness of the surface does not seem to have a significant effect 
%on the overall propagation of the emission out of the ice, at least in
%terms of the overall field strength.

In summary, Askaryan's hypothesis has now been confirmed in detail 
by laboratory experiments for
virtually all of the dielectrics (ice, salt, sand--the latter 
approximating the Lunar regolith) that Askaryan envisioned as the
best media in which to exploit the coherent radio Cherenkov emission
from high energy particle showers. Askaryan's intent was to illuminate
a methodology by which low fluxes of ultra-high energy particles could be made
observable through exploitation of huge volumes of natural materials.
With the recent sharpening of predictions for the fluxes of 
ultra-high energy neutrinos, and the growth in the number of experiments that
make use of it, we expect that Askaryan's hope will be soon
fulfilled.

This work has been supported by the National Aeronautics and Space
Administration and the Department of Energy Office of Science High Energy
Physics Division. We thank the SLAC Experimental Facilities Department and
the Columbia Scientific Balloon Facility for their invaluable support.

\end{document}